\documentstyle[eqsecnum,floats,preprint,aps]{revtex}
\tighten

\begin{document}

\newcommand{\vp}{\varphi}
\newcommand{\CP}{${\cal CP}$}
\newcommand{\C}{${\cal C}$}
\newcommand{\Parity}{${\cal P}$}
\newcommand{\B}{${\cal B}$}
\newcommand{\be}{\begin{equation}}
\newcommand{\ee}{\end{equation}}
\newcommand{\bea}{\begin{eqnarray}}
\newcommand{\eea}{\end{eqnarray}}
\newcommand{\PSbox}[3]{\mbox{\rule{0in}{#3}\includegraphics{#1}\hspace{#2}}}

\newcommand{\bref}[1]{(\ref{#1})}
\newcommand{\Lag}{{\cal L}}
\newcommand{\th}{\theta}
\newcommand{\thb}{\bar{\theta}}
\newcommand{\al}{\alpha}
\newcommand{\ald}{{\dot{\alpha}}}
\newcommand{\sgvp}{\sigma^\vp}
\newcommand{\sgr}{\sigma^r}
\newcommand{\sgz}{\sigma^z}
\newcommand{\la}{\lambda}
\newcommand{\lab}{\bar{\lambda}}
\newcommand{\psb}{\bar{\psi}}
\newcommand{\phb}{\bar{\phi}}
\newcommand{\xib}{\bar{\xi}}
\newcommand{\chb}{\bar{\chi}}
\newcommand{\dmu}{{\partial_\mu}}
\newcommand{\dth}{{\partial_\vp}}
\newcommand{\dr}{{\partial_r}}

\newcommand{\Ch}{\Phi}
\newcommand{\Chc}{\tilde{\Phi}}
\newcommand{\bfCh}{{\bf \Phi}}
\newcommand{\bfChc}{{\bf \tilde{\Phi}}}
\newcommand{\Sca}{S}
\newcommand{\Scac}{{\tilde{S}}}
\newcommand{\Scao}{{S_0}}

\newcommand{\phc}{\tilde{\phi}}
\newcommand{\bfphi}{\mbox{\boldmath $\phi$}}
\newcommand{\bfphc}{\mbox{\boldmath $\phc$}}
\newcommand{\psc}{\tilde{\psi}}
\newcommand{\bfpsi}{\mbox{\boldmath $\psi$}}
\newcommand{\bfpsc}{\mbox{\boldmath $\psc$}}
\newcommand{\Fc}{\tilde{F}}

\newcommand{\sca}{s}
\newcommand{\scac}{\tilde{s}}
\newcommand{\scao}{s_0}
\newcommand{\scf}{\omega}
\newcommand{\scfx}[1]{\omega_{({#1})}}
\newcommand{\scfc}{\tilde{\omega}}
\newcommand{\scfo}{\omega_0}

\newcommand{\e}[1]{{\bf e}_{{#1}}}
\newcommand{\ept}{e^{i\vp}}
\newcommand{\emt}{e^{-i\vp}}
\newcommand{\qp}{q^+}
\newcommand{\qm}{q^-}
\newcommand{\qA}{q^A}
\newcommand{\qB}{q^B}

\newcommand{\bmat}[1]{\left(\begin{array}{{#1}}} 
\newcommand{\emat}{\end{array}\right)}

\title{Cosmic Strings, Zero Modes and SUSY breaking in Nonabelian
$N=1$ Gauge Theories}

\author{Stephen C. Davis$^{1}$\footnote[1]{S.C.Davis@damtp.cam.ac.uk}, 
Anne-Christine Davis$^{1}$\footnote[2]{A.C.Davis@damtp.cam.ac.uk} and
Mark Trodden$^{2}$\footnote[3]{trodden@theory1.physics.cwru.edu.}}

\address{~\\$^1$Department of Applied Mathematics and Theoretical Physics \\
University of Cambridge, CB3 9EW, UK.}

\address{~\\$^2$Particle Astrophysics Theory Group\\ 
Department of Physics \\
Case Western Reserve University \\
10900 Euclid Avenue \\
Cleveland, OH 44106-7079, USA.}

\maketitle

\begin{abstract}

We investigate the microphysics of cosmic strings in
Nonabelian gauge theories with $N=1$ supersymmetry. 
We give the vortex solutions in a specific example and demonstrate that 
fermionic superconductivity arises because of the couplings and
interactions dictated by supersymmetry. We then use supersymmetry 
transformations to obtain the relevant fermionic zero modes and investigate 
the role of soft supersymmetry breaking on the existence and properties of 
the superconducting strings. 

\end{abstract}

\setcounter{page}{0}
\thispagestyle{empty}

\vfill

\noindent DAMTP-97-116\hfill Submitted to {\it Physics Review} {\bf D}

\noindent CWRU-P21-97

\hfill Typeset in REV\TeX

\eject

\vfill

\eject

\baselineskip 24pt plus 2pt minus 2pt

\section{Introduction}
In recent years, supersymmetry (SUSY) has become increasingly favoured 
as the theoretical structure underlying fundamental particle interactions.
In this light it is natural to investigate possible cosmological implications
of SUSY. 

In a recent paper~\cite{DDT 97} we discussed the effect of SUSY on
the microphysics of simple cosmic string solutions of abelian field theories.
In particular we developed and applied the technique of SUSY transformations 
to investigate the form of fermionic zero modes, required by SUSY, which
lead to cosmic string superconductivity. In the present work we extend our
original ideas to a more general class of field theories, namely those with
a nonabelian gauge group. Since nonabelian gauge theories underlie modern
particle physics, and in particular, unified field theories, this class
of theories is a realistic toy model for grand unified theories (GUT). The 
particular example we examine, 
$SU(2)\times U(1)\rightarrow U(1)\times Z_2$, admits two types
of string solution, one abelian and the other nonabelian. This model has a
very similar structure to $SO(10)$ and should provide insight into cosmic
strings in SUSY GUTs. Most of the features exhibited by this theory will
also appear in larger non-abelian theories. We apply the
technique of SUSY transformations to the nonabelian case and extract the 
behaviour of the zero modes as functions of the background string fields. We
then compare the results to those obtained in Ref.~\cite{DDT 97} for 
abelian strings.

Since SUSY is clearly broken in the universe today, it is important to know
how the SUSY zero modes behave when soft-SUSY breaking occurs in the universe.
We investigate this for both the abelian and nonabelian strings by explicitly 
introducing soft-SUSY breaking terms into the theory. The result is that
all the zero modes are destroyed in almost all the theories; the
exception being when a Fayet-Iliopoulos term is used to break the gauge
symmetry in an abelian model. We briefly comment on the physical reasons
for this and show how the effect may be seen through the breakdown of an
appropriate index theorem.

These results have cosmological significance since fermion zero modes on the
string can be excited, causing a current to flow along the string. The
string then behaves as a perfect conductor. The existence of charge carriers
changes the cosmology of cosmic strings. In particular, they can stabilise
string loops, resulting in the production of so-called 
vortons~\cite{oldvortons}.
Such vortons can dominate the energy density of the Universe, and have been
used to constrain GUT models with current-carrying strings~\cite{vortons}.
However, if the zero modes are destroyed at the SUSY breaking energy scale, 
then the current, and hence vortons, will dissipate. Thus, the underlying 
theory may be cosmologically viable.

The plan of this paper is as follows. In section 2 we construct a simple
supersymmetric model based on the group $SU(2)\times U(1)$ and display 
the abelian and non-abelian string solutions. In section~3, we briefly
review a powerful index theorem for finding fermion zero modes in a general
theory. We then use SUSY transformations to obtain the zero modes in terms
of the background string fields. Soft SUSY breaking terms are introduced in 
section 4 and their 
effect on the zero modes is analysed using the results of the index theorem.
In this section we consider both the string solutions for the $SU(2)\times
U(1)$ model and also for the $U(1)$ theory discussed in our previous 
paper~\cite{DDT 97}. For the non-abelian string
the SUSY breaking terms destroy the zero modes, while for the other string 
solutions the situation is more complicated.  
 
\section{A Simple Model: $SU(2)\times U(1)$}
There exist many nonabelian theories with breaking schemes giving rise to
cosmic strings. In general both abelian and nonabelian strings can be 
produced in such a process, depending on which part of the vacuum manifold
is involved in the winding. 

In this section we consider a simple example in which the gauge group 
$SU(2) \times U(1)$ is spontaneously broken down to the group 
$U(1) \times Z_2$ via the superpotential

\be 
W = \mu_1 \Scao (\bfCh \cdot \bfChc - \eta^2) 
    + \mu_2 (\Sca \bfCh^T \Lambda \bfCh + \Scac \bfChc^T \Lambda \bfChc) \ .
\ee
The chiral superfields $\Ch_i(\phi_i,\psi_i,F_i)$ and 
$\Chc_i(\phc_i,\psc_i,\Fc_i)$ are $SU(2)$ triplets with $U(1)$ 
charges $\pm1$ respectively. the other chiral superfields,
$\Scao(\scao, \scfo, F_\Scao)$, $\Sca(\sca, \scf, F_\Sca)$ and 
$\Scac(\scac, \scfc, F_\Scac)$, are $SU(2)$ scalars with $U(1)$ charges 0, 
$-2$ and $+2$ respectively. Finally, defining $T^4 = \sqrt{2/3} I$, 
the vector supermultiplets are $V^a(A^a_\mu,\la^a,D^a)$, $a=1 \ldots 4$.
Since the constant matrix $\Lambda$ satisfies 
$\Lambda T^i = -(T^i)^\ast \Lambda$ 
($i=1 \ldots 3$) and the $SU(2)$ gauge transformations are 
$\delta \bfCh = i T^a n^a \bfCh$ and 
$\delta \bfChc = -i T^{a\ast} n^a \bfChc$, the superpotential is gauge 
invariant.

The scalar potential, derived in the standard manner~\cite{susybook},  is then

\bea
U &=& \mu_1^2 | \bfphi \cdot \bfphc - \eta^2 |^2
    + \mu_2^2 | 2 \phi_1 \phi_3  - \phi_2^2 |^2
    + \mu_2^2 | 2 \phc_1 \phc_3  - \phc_2^2 |^2
\nonumber \\ &&{}
    + | \mu_1 \scao \bfphc + 2 \mu_2 \sca \Lambda \bfphi |^2 
    + | \mu_1 \scao \bfphi + 2 \mu_2 \scac \Lambda \bfphc |^2
\nonumber \\ &&{} 
 + e^2 |(\phi_1 + \phi_3)\phi_2^\ast - (\phc_1 + \phc_3)\phc_2^\ast|^2
\\ &&{}
 + \frac{e^2}{2} (|\phi_1|^2 - |\phi_3|^2 - |\phc_1|^2 +|\phc_3|^2 )^2 
\nonumber \\ &&{}
 + \frac{e^2}{3} (|\bfphi|^2 - |\bfphc|^2 - 2|\sca|^2 + 2|\scac|^2 )^2
\ . \nonumber 
\eea
This is minimised when all fields are zero except $\phi_1 = \phc_1 =
\eta$, or at any (broken) gauge transformation of this. We note also
that the theory has a local minimum with $\bfphi = \bfphc = 0$ and
that this structure can give rise to hybrid
inflation~\cite{inflation}. This is true even for the abelian theory
described in~\cite{DDT 97}. In both cases inflation ends with defect
formation. 

As we mentioned above, there are abelian and nonabelian string
solutions to this theory. The abelian solution is obtained from the ansatz

\bea
\phi_1 = \phc^\ast_1 &=& \eta f(r) \ept \ ,\\
A_\vp   &=& \frac{a(r)}{er} \sqrt{\frac{3}{5}} T^G \ ,\\
F_\Scao &=& \mu_1 \eta^2 (1 - f(r)^2)  \ ,
\eea
where $T^G = \sqrt{\frac{3}{5}}T^3 + \sqrt{\frac{2}{5}}T^4$. All other
fields are zero and the profile functions
functions $a(r)$ and $f(r)$ obey the boundary conditions $f(0)=a(0)=0$ and 
$\lim_{r\rightarrow\infty}f(r)=\lim_{r\rightarrow \infty} a(r)=1$.

The nonabelian solution is obtained from the ansatz

\bea
\bfphi = \bfphc^\ast &=& 
		\eta \left\{ \frac{1}{2} (\ept \e{+} + \emt \e{-}) f(r) 
			+ \frac{1}{\sqrt{2}} \e{0} g(r) \right\}\ , \\
A_\vp   &=& \frac{a(r)}{er} T^1 \ ,  \\
F_\Scao &=& \frac{1}{2} \mu_1 \eta^2 [2 - f(r)^2 - g(r)^2] \ , \\
F_\Sca = F_\Scac &=& \frac{1}{2} \mu_2 \eta^2 [f(r)^2 - g(r)^2] \ ,
\eea  
where $\e{k}$ are unit vectors obeying $T^1 \e{k} = k\e{k}$. In this
case $g(0)$ is finite, $\lim_{r\rightarrow \infty} g(r)=1$ and $f(r)$ and 
$a(r)$ obey the same boundary conditions as in the abelian case. 

Note that $f$, $g$ and $a$ are solutions to simple coupled second
order ordinary differential equations. Their forms can be obtained
numerically and are well-known~\cite{Ma}.

\section{SUSY Transformations and Zero-Modes}

The string solutions obtained above have all the fermion fields set to zero.
In this section we investigate what happens when these fields are excited in
the background of the cosmic string. 

We can find the number of zero modes with an appropriate index
theorem~\cite{Index}. To apply the theorem, the
fermion fields must be expressed as eigenstates of the string
generator. The resulting mass matrix is then split up into irreducible
parts, and the theorem applied to each part separately.

For most irreducible mass matrices, the eigenvalues of the fermion
fields are split into positive ($\qp_i$) and negative ($\qm_j$)
eigenvalues, and then ordered. If there are $n_+$ positive and $n_-$
negative eigenvalues, then 
$\qp_1 \geq \qp_2 \geq \ldots \geq \qp_{n_+} > 0$ and 
$\qm_1 \leq \qm_2 \leq \ldots \leq \qm_{n_-} < 0$. Any zeros are
ignored. The numbers of complex left and right moving zero modes are then 
given by

\bea
N_L &=& \sum^{\min(n_-,n_+)}_{j=1} [\qp_j + \qm_j]_+ 
	+ \sum^{n_+}_{j=n_- +1} \qp_j \ , \\ \nonumber
N_R &=& \sum^{\min(n_-,n_+)}_{j=1} [-\qp_j -\qm_j]_+ 
	+ \sum^{n_-}_{j=n_+ +1} - \qm_j \ .
\label{index1}
\eea
where $[x]_+ = x$ if $x>0$, and 0 otherwise.

The exception to this is when the mass matrix can be put into the form

\be
\bmat{cc} 0 & A \\ B & 0 \emat \ ,
\ee
where $A$ and $B$ are $n \times n$ matrices. If the fermion states are
put in a corresponding form, then in this case the eigenvalues
are split up according whether their eigenstates occupy the first or
second half of the $2n$ component fermion vector. The two sets of
eigenvalues are then ordered so that
$\qA_1 \geq \qA_2 \geq \ldots \geq \qA_n$ and 
$\qB_1 \leq \qB_2 \leq \ldots \leq \qB_n$. The resulting number of zero modes
is 

\bea
N_L &=& \sum^{n}_{j=1} [\qA_j + \qB_j]_+ \ , \\ \nonumber
N_R &=& \sum^{n}_{j=1} [-\qA_j -\qB_j]_+ \ .
\label{index2}
\eea

The derivation of this theorem assumed that none of the fermion fields
involved were massless at large $r$. In fact the results still hold
for massless fields if all the eigenvalues are whole integers.

We therefore know that there must exist fermion zero modes on the string. 
Rather than attempting to solve the difficult fermion equations of
motion to obtain  these solutions, we shall use a different
technique~\cite{DDT 97} which exploits the power of SUSY to obtain
some of the solutions.

\subsection{The Abelian String}
The relevant part of the Lagrangian is the Yukawa sector which is entirely
determined by supersymmetry. In the abelian case, the nonzero Yukawa 
couplings are

\bea 
\Lag_Y &=& -\left[
	 \mu_1 (\ept \psc_1 + \emt \psi_1) \scfo
 	 +\sqrt{\frac{5}{6}}ie (\emt \psi_1 - \ept \psc_1) \la^G 
       \right. \nonumber \\ && \left.
	 + 2\mu_2 (\ept \psi_3 \scf + \emt \psc_3 \scfc) 
       \right. \nonumber \\ && \left.
	 + \frac{ie}{\sqrt{2}} (\emt \la_+ \psi_2 + \ept \la_- \psc_2)
       \right] \eta f(r) \ , 
\label{abYuk}
\eea
where $\la_\pm = (\la^1 \mp i\la^2)/\sqrt{2}$. With respect to the
string generator, the only fields with non-zero eigenvalues are
$\psi_1$ and $\la_+$ (eigenvalue 1), and $\psc_1$ and $\la_-$
(eigenvalue $-1$). The Yukawa Lagrangian can be split up into 5 independent
parts. Applying \bref{index2} to these reveals that there are a total of three
left moving and three right moving complex zero-modes.

Now, following the techniques of~\cite{DDT 97}, we perform an infinitesimal
SUSY transformation (with Grassmann parameter $\xi_{\alpha}$) followed by a 
gauge transformation to return to Wess-Zumino gauge. The (bosonic) string 
fields
all transform quadratically and so are unchanged to first order. However, the
fermions transform linearly and, in terms of the background string fields
it is possible to find two complex (or 4 real) fermion zero mode
solutions given by

\bea
\scfo &=& \sqrt{2} \mu_1 \eta^2 (1 - f^2) \xi \ , \\
\la^G &=& -i\sqrt{\frac{3}{5}} \frac{a}{er} \sgz \xi \ , \\
\psi_1 &=& i\sqrt{2} \eta \ept
	\left[ f' \sgr + i\frac{(1-a)}{r}f \sgvp \right] \xib \ , \\
\psc_1 &=& i\sqrt{2} \eta \emt
	\left[ f' \sgr - i\frac{(1-a)}{r}f \sgvp \right] \xib \ .
\label{abzeromodes}
\eea
Setting either component of $\xi$ to zero gives one of the zero modes,  
one of which is left moving and the other right moving.

\subsection{The Nonabelian String}
In the nonabelian case it is convienient to split $\bfpsi$, $\bfpsc$
and $\la^i$ into eigenvectors of $T^1$, and label them by their
eigenvalues.  Defining $\chi_i = (\psi_i + \psc_i)/\sqrt{2}$, 
$\zeta_i = (\psi_i - \psc_i)/\sqrt{2}$, and 
$\scfx{\pm} = (\scf \pm \scfc)/\sqrt{2}$, the nonzero Yukawa couplings are
\bea 
\Lag_Y &=& -\mu_1\eta\left[ \chi_0 g(r) + 
	 \frac{1}{\sqrt{2}}(\chi_+\emt + \chi_-\ept)f(r)\right] \scfo
       \nonumber \\ &&
	 -\mu_2 \eta \left[-\sqrt{2}\chi_0 g(r)
	 + (\chi_+\emt + \chi_-\ept)f(r)\right]\scfx{+}
       \nonumber \\ &&
	 - \frac{ie\eta}{2} (\chi_+\emt - \chi_-\ept)f(r) \la^0
       \nonumber \\ &&
	 - \mu_2\eta \left[-\sqrt{2}\zeta_0 g(r)
	 + (\zeta_+\emt +
	 \zeta_-\ept)f(r)\right]\scfx{-}
       \nonumber \\ &&
	 + \frac{e\eta}{2}\left[\sqrt{2} (\zeta_-\la^+ - \zeta+\la^-) g(r)
	 + \zeta_0(\la^+ \emt - \la^- \ept)f(r) \right]
       \nonumber \\ &&
	 - \frac{ie\eta}{\sqrt{6}}\left[ \sqrt{2} \zeta_0 g(r)
	 + (\zeta_+ \emt + \zeta_- \ept)f(r)\right] \la^4 \ .
\label{nabLagY}
\eea
In this case $\chi_\pm$, $\zeta_\pm$ and $\la^\pm$ have eigenvalues
$\pm 1$.  Applying \bref{index2} to the two irreducible parts of
\bref{nabLagY} shows there are just two complex zero modes, moving in
opposite directions.

Once again performing an infinitesimal SUSY transformation and a (nonabelian)
gauge transformation we obtain the two complex zero modes

\bea
\scfo &=& \frac{1}{\sqrt{2}} \mu_1 \eta^2 (2 - g^2 - f^2) \xi \ , \\
\scfx{+} &=& \mu_2 \eta^2 (g^2 - f^2) \xi \ , \\
\la^0 &=& -i \frac{a}{er} \sgz \xi \ , \\
\chi_+ &=& i \eta \ept
	\left( f' \sgr + i\frac{1-a}{r}f \sgvp \right) \xib \ , \\
\chi_- &=& i \eta\emt
	\left( f' \sgr - i\frac{1-a}{r}f \sgvp \right) \xib \ , \\
\chi_0 &=& i \sqrt{2}\eta g' \sgr \xib \ . 
\label{nabzeromodes}
\eea
Thus in this case there are no zero modes beyond those implied by
SUSY, in contrast to the abelian case. This is related to the fact
that there are components of the Higgs fields that do not wind in the
nonabelian case. 

\section{Soft Susy Breaking}
Perhaps the most attractive feature of SUSY arises from the nonrenormalization
theorems. These ensure that quadratic divergences are absent in SUSY theories
and so protect any tree-level hierarchy of scales from receiving quantum
corrections.

When SUSY is broken, as it must be at a scale of around $1$TeV, it is crucial
that these quadratic divergences remain absent from the theory. This is 
achieved by adding only {\it soft}\ SUSY breaking terms to the model. Such 
terms are defined as being those which are noninvariant under SUSY 
transformations but which do not induce quadratic divergences.

In a general model, one may obtain soft SUSY breaking terms by the 
following prescription.

\begin{enumerate}
\item Add arbitrary mass terms for all scalar particles to the scalar 
potential.

\item Add all trilinear scalar terms in the superpotential, plus
their hermitian conjugates, to the scalar potential with arbitrary coupling.

\item Add mass terms for the gauginos to the Lagrangian density.
\end{enumerate}

Since the techniques we have used are strictly valid 
only when SUSY is exact, it is necessary to investigate the effect of these 
soft terms on the fermionic zero modes we have identified.

As we have already commented, the existence of the zero modes can be seen as 
a consequence of an index theorem~\cite{Index}. The index is
insensitive to the size and exact form of the Yukawa couplings, as
long as they are regular for small $r$, and tend to a constant at
large $r$. In fact, the existence of zero modes relies only on
the existence of the appropriate Yukawa couplings and that they have the 
correct $\vp$-dependence. Thus there can only be a change in the number of zero
modes if the soft breaking terms induce specific new Yukawa couplings
in the theory and it is this that we must check for. Further, it was
conjectured in~\cite{Index} that the destruction of a zero mode occurs only 
when the relevant fermion mixes with another massless fermion. 

We have examined each of our theories, both from this paper and 
Ref.~\cite{DDT 97}, with respect to this criterion and list the results 
below.

\subsection{$U(1)$ Abelian Models}

\subsubsection{Theory-F}

In Ref.~\cite{DDT 97} we referred to an abelian theory in which the gauge
symmetry is broken via an $F$-term as ``theory-$F$''. The corresponding
superpotential was
\be
W = \mu \Phi_0 (\Phi_+ \Phi_- - \eta^2) \ .
\ee
The trilinear and mass terms that arise from soft SUSY breaking
in this theory are
\be
m_0^2 |\phi_0|^2 + m_-^2 |\phi_-|^2 + m_+^2 |\phi_+|^2 
 + \mu M \phi_0\phi_+\phi_-
\ee
The derivative of the scalar potential with respect to $\phi_0^\ast$ becomes
\be
\phi_0 (\mu^2|\phi_+|^2 + \mu^2|\phi_-|^2 + m_0^2) + \mu M(\phi_+ \phi_-)^\ast
\ee
This will be zero at a minimum, and so $\phi_0 \ne 0$ only if $M \ne 0$.

New Higgs mass terms will alter the values of $\phi_+$ and $\phi_-$ slightly, 
but will not produce any new Yukawa terms. Thus these soft
SUSY-breaking terms have no effect on the existence of the zero modes.

However, the presence of the trilinear term gives $\phi_0$ a non-zero 
expectation value, which gives a Yukawa term coupling the $\psi_+$ and $\psi_-$
fields. This destroys all the zero modes in the theory since the left
and right moving zero modes mix. 

For completeness note that a gaugino mass term also mixes the left and right 
zero modes, aiding in their destruction.
  
In terms of the index theorem, the change in the number of zero modes
arises because \bref{index1} applies after the SUSY breaking, while
\bref{index2} applied before it. Although the fermion eigenvalues do
not change, the expression relating them to the zero modes does.

\subsubsection{Theory-D}

The $U(1)$ theory with gauge symmetry broken via a Fayet-Iliopoulos term and 
no superpotential is simpler to analyse~\cite{DDT 97}. New Higgs mass terms 
have no effect, as in the above  case, and there are no trilinear
terms. Further, although the gaugino mass  terms affect the form of
the zero mode solutions, they do not affect their  existence, and so,
in theory-$D$, the zero modes remain even after SUSY breaking. 

\subsection{$SU(2)\times U(1)$ Model}

\subsubsection{Abelian Strings}

The effect of soft SUSY breaking terms on the zero modes which were
found analytically in \bref{abzeromodes} is identical to the equivalent $U(1)$
theory. Thus, all SUSY zero modes are destroyed in this case. In this
larger theory, there are also other, non-SUSY, zero modes.
Not all of these are destroyed by a gaugino mass term, as some
do not involve the gaugino fields. However, if the trilinear terms give
$\scao$ a non-zero vacuum expectation value, all the zero modes are
destroyed. The extra Yukawa terms mean there are fewer irreducible
parts to the fermion mass matrices. This results in more terms
cancelling in \bref{index2}, reducing the number of zero modes. As in
the other cases, the physical reason behind the destruction of the
zero modes is that left and right movers mix. 

\subsubsection{Nonabelian Strings}

As in the other cases above, nonzero gaugino mass or trilinear terms
destroy the zero modes that were found with SUSY transformations 
\bref{nabzeromodes}. Similarly \bref{index1} is required instead of
\bref{index2}, implying that the left and right moving modes mix. For 
nonabelian strings these are the only zero modes and so none remain after 
SUSY breaking.

\section{Comments and Conclusions}
We have examined the microphysics of abelian and nonabelian cosmic string 
solutions to the $SO(10)$ inspired supersymmetric $SU(2)\times U(1)$ model. 
By performing infinitesimal SUSY transformations on the background string
fields we have obtained the form of the fermionic zero modes responsible for
cosmic string conductivity. These solutions may be compared to those
found in our earlier work~\cite{DDT 97} on abelian defects. 

Our results mean that fermion zero modes are always present around cosmic 
strings in SUSY. We conjecture that in theories with $F$-term gauge symmetry 
breaking, the zero modes given by SUSY always occur in pairs, one left and one
right moving. It also seems likely that such theories always have
hybrid inflation.

Further, in the abelian case there were additional zero
modes that were not a consequence of supersymmetry. We expect that
similar extra zero modes will be present in a larger theory, even in
the nonabelian case. 

We have also analysed the effect of soft SUSY breaking on the existence 
of fermionic zero modes. The $SU(2)\times U(1)$ model and two simple abelian
models were examined. In all cases Higgs mass terms did not affect the
existence of the zero modes. In the theories with $F$-term symmetry
breaking, gaugino mass terms destroyed all zero modes which involved
gauginos and trilinear terms created extra Yukawa couplings which
destroyed all the zero modes present. 
In the abelian theory with $D$-term symmetry breaking, the zero modes
were unaffected by the SUSY breaking terms. It was conjectured
in~\cite{Index} that zero modes would only disappear when they mixed
with another massless fermion field and this is consistent with the
results obtained in this chapter.  If the remaining zero modes survive
subsequent  phase transitions, then stable vortons could result. Such
vortons would dominate the energy density of the universe, rendering
the underlying GUT cosmologically problematic.

Therefore, although SUSY breaking may alleviate the cosmological disasters 
faced by superconducting cosmic strings~\cite{vortons}, there are classes of
string solution for which zero modes remain even after SUSY breaking. It
remains to analyse all the phase transitions undergone by specific SUSY
GUT models to see whether or not fermion zero modes survive down to the
present time. If the zero modes do not survive SUSY breaking, the
universe could experience a period of vorton domination beforehand,
and then reheat and evolve as normal afterwards.

If the zero modes do occur in pairs (one left and one right moving) in
$F$-term gauge symmetry breaking, it is possible that they could 
scatter off each other~\cite{BarrMatheson}. This would cause the current to 
decay, and could stop vorton domination.

There is the possibility that even if zero modes
are destroyed they become low-lying bound states. Such bound states may
still be able to carry a persistent current. If this is the case, even 
such theories may not be safe cosmologically. Work on this is under
investigation.

It may also be possible to extend our analysis of the effect of SUSY breaking 
on the bosonic fields. The resulting potential is very complex, even in the
abelian case. However it may be possible to use some sort of
approximation, or numerical solution. The change in potential also
affects the hybrid inflation which occurs in the model, although at this
stage it is not clear how.

\acknowledgments

This work is supported in part by PPARC and the E.U. under the HCM program 
(CHRX-CT94-0423) (S.C.D. and A.C.D.), and by Trinity College Cambridge 
(S.C.D.). M.T. is supported by the U.S. Department of Energy (D.O.E.), the 
National Science foundation (N.S.F.) and by funds provided by Case Western 
Reserve University.

\end{document}